\documentclass[sigconf]{acmart}

\usepackage{graphicx} %
\usepackage{natbib}  %
\usepackage{caption} %
\usepackage{algorithm}
\usepackage{listings}
\usepackage{algorithmic}
\usepackage{amsmath}
\usepackage{booktabs}
\usepackage{multirow}
\usepackage[outdir=./]{epstopdf}
\usepackage{enumitem}
\usepackage{caption}
\usepackage{subcaption}
\usepackage{newfloat}
\usepackage{graphicx}
\usepackage{svg} % svg
\usepackage[normalem]{ulem}

\DeclareMathOperator*{\argmax}{arg\,max}
\newlist{todolist}{itemize}{2}
\setlist[todolist]{label=$\square$}
\usepackage{pifont}

\newtheorem{mdefinition}{Definition}

\newcommand{\para}[1]{{\vspace{4pt} \bf \noindent #1 \hspace{0pt}}}

\usepackage{array}
\newcolumntype{L}[1]{>{\raggedright\let\newline\\\arraybackslash\hspace{0pt}}m{#1}}
\newcolumntype{C}[1]{>{\centering\let\newline  \\\arraybackslash\hspace{0pt}}m{#1}}
\newcolumntype{R}[1]{>{\raggedleft\let\newline \\\arraybackslash\hspace{0pt}}m{#1}}

\AtBeginDocument{%
  \providecommand\BibTeX{{%
    \normalfont B\kern-0.5em{\scshape i\kern-0.25em b}\kern-0.8em\TeX}}
}

\copyrightyear{2023}
\acmYear{2023}
\setcopyright{acmlicensed}
\acmConference[KDD '23] {Proceedings of the 29th ACM SIGKDD Conference on Knowledge Discovery and Data Mining}{August 6--10, 2023}{Long Beach, CA, USA.}
\acmBooktitle{Proceedings of the 29th ACM SIGKDD Conference on Knowledge Discovery and Data Mining (KDD '23), August 6--10, 2023, Long Beach, CA, USA}
\acmPrice{15.00}
\acmISBN{979-8-4007-0103-0/23/08}
\acmDOI{10.1145/3580305.3599322}

\begin{document}

\title{Efficient and Joint Hyperparameter and Architecture Search for Collaborative Filtering}

\author{Yan Wen}
\orcid{0009-0002-6425-5056}
\affiliation{
  \institution{Department of Electronic Engineering}
  \institution{Beijing National Research Center for Information Science and Technology}
  \institution{Tsinghua University}
  \city{Beijing}
  \country{China}
}
\email{wenyan0531@gmail.com}

\author{Chen Gao}
\orcid{0000-0002-7561-5646}
\authornote{Corresponding author.}
\affiliation{
	\institution{Department of Electronic Engineering}
	\institution{Beijing National Research Center for Information Science and Technology}
   	\institution{Tsinghua University}
  \city{Beijing}
  \country{China}
}
\email{chgao96@gmail.com}

\author{Lingling Yi}
\orcid{0000-0001-8809-7676}
\affiliation{
  \institution{Tencent Inc.}
  \city{Shenzhen}
  \country{China}
}
\email{chrisyi@tencent.com}

\author{Liwei Qiu}
\orcid{0009-0006-9742-6762}
\affiliation{
	\institution{Tencent Inc.}
	\city{Shenzhen}
	\country{China}
}
\email{drolcaqiu@tencent.com}

\author{Yaqing Wang}
\orcid{0000-0003-1457-1114}
\affiliation{
	\institution{Baidu Inc.}
	\city{Beijing}
	\country{China}
}
\email{wangyaqing01@baidu.com}

\author{Yong Li}
\orcid{0000-0001-5617-1659}
%\authornote{Corresponding author.}
\affiliation{
	\institution{Department of Electronic Engineering}
	\institution{Beijing National Research Center for Information Science and Technology}
	\institution{Tsinghua University}
	\city{Beijing}
	\country{China}
}
\email{liyong07@tsinghua.edu.cn}

\renewcommand{\shortauthors}{Yan Wen et al.}

\begin{abstract}

Automated Machine Learning (AutoML) techniques have recently been introduced to design Collaborative Filtering (CF) models in a data-specific manner.
However, existing works either search architectures or hyperparameters while ignoring the fact they are intrinsically related and should be considered together.
This motivates us to consider a joint hyperparameter and architecture search method to design CF models.
However, this is not easy because of the large search space and high evaluation cost.
To solve these challenges, we reduce the space by screening out usefulness hyperparameter choices through a comprehensive understanding of individual hyperparameters.
Next, we propose a two-stage search algorithm to find proper configurations from the reduced space. 
In the first stage, we leverage knowledge from subsampled datasets to reduce evaluation costs;
in the second stage, we efficiently fine-tune top candidate models on the whole dataset.
Extensive experiments on real-world datasets show  better performance can be achieved compared with both hand-designed and previous searched models.
Besides, ablation and case studies demonstrate the effectiveness of our search framework.

\end{abstract}


% \ccsdesc[500]{Data Science~Recommender Systems}


%\settopmatter{printfolios=true} % add pages
\maketitle

\section{Introduction}
\label{sec::intro}

Collaborative Filtering (CF) is the most widely used approach for Recommender Systems~\cite{sarwar2001item,koren2008factorization,he2017neural,kabbur2013fism}, aiming at calculating the similarity of users and items to recommend new items to potential users. 
They mainly use Neural Networks to build models for users and items, simulating the interaction procedure and predict the preferences of users for items.
Recent works also built CF models based on Graph Neural Networks (GNNs)~\cite{gao2023survey,ying2018graph, wang2019neural, he2020lightgcn}.
While CF models may have different performance on different scenes~\cite{dacrema2019we}, recent works~\cite{zheng2022automlrecsys,chen2022automated} have begun to apply Automated Machine Learning (AutoML) to search data-specific CF models.
Previous works, including SIF~\cite{yao2020searching}, AutoCF~\cite{gao2021efficient} and~\cite{wang2022profiling}, applied Neural Architecture Search (NAS) on CF tasks.
They split the architectures of CF models into several parts and searched each part on architecture space. 

However, most of these methods focus on NAS in architecture space, 
only considering hyperparameters as fixed settings and therefore omitting the dependencies among them. 
A CF model can be decided by a given architecture and a configuration of hyperparameters. 
Especially in the task of searching best CF models, hyperparameter choice can affect search efficiency and the evaluation performance an architecture can receive on a given dataset. 
Recent methods mainly focus on model search, ignoring the important role of hyperparameters.
For instance, SIF focuses on interaction function, while it uses grid search on hyperparameters space.
AutoCF does not use hyperparameter tuning on each architecture, which may make the searched model sub-optimal since proper hyperparameters vary for different architectures.
\cite{wang2022profiling} includes GNN models in the architecture space, but the search and evaluation cost for architectures may be high.

We find that these works only focus on either hyperparameters or fixed parts in CF architecture, neglecting the relation between architecture and hyperparameters.
If architecture cannot be evaluated with proper hyperparameters, 
a sub-optimal model may be searched, possibly causing a reduction in performance. 
To summarize, there exists a strong dependency between hyperparameters and architectures.
That is, the hyperparameters of a model are based on the design of architectures, 
and the choices of hyperparameters also affect the best performance an architecture may approach. 
We suppose that hyperparameters can be adaptively changed when the architecture change in CF tasks, 
so we consider that the CF search problem can be defined on a joint space of hyperparameters and architectures.
Therefore, the CF problem can be modeled as a joint search problem on architecture and hyperparameter space.
While hyperparameters and architectures both influences the cost and efficiency of CF search task, 
there exist challenges for joint search problems:
(1) Since the joint search space is designed to include both hyperparameters and architectures, the joint search problem has a large search space, 
which may make it more difficult to find the proper configuration of hyperparameters and architectures;
(2) In a joint search problem, since getting better performance on a given architecture requires determining its hyperparameters, the evaluation cost may be more expensive.

We propose a general framework, which can optimize CF architectures and their hyperparameters at the same time in search procedure. 
The framework of our method is shown in Figure~\ref{fig:search_two_stage}, consisting of two stages.
Prior to searching hyperparameters and architectures,  we have a full understanding of the search space and reduce hyperparameter space to improve efficiency.
Specifically, we reduced the hyperparameter space by their performance ranking on CF tasks from different datasets.
We also propose a frequency-based sampling strategy on the user-item matrix for fast evaluation.
%Considering the CF dataset as a user-item bipartite graph, we sample users and items with higher frequency of interaction.
In first stage, we search and evaluate models on reduced space and subsampled datasets, and we jointly search architecture and hyperparameters with a surrogate model.
In second stage, we propose a knowledge transfer-based evaluation strategy to leverage surrogate model to larger datasets.
Then we evaluate the model with transferred knowledge and jointly search the hyperparameters and architectures to find the best choice in original dataset. 

Overall, we make the following important contributions:
%\begin{itemize}[leftmargin=*]
\begin{itemize}[noitemsep, topsep=2pt, leftmargin=8pt]
	\item We propose an approach that can jointly search CF architectures and hyperparameters to 
	get performance from different datasets.
	
	\item We propose a two-stage search algorithm to efficiently optimize the problem. 
	The algorithm is based on a full understanding of search space and transfer ability between datasets.
	It can jointly update CF architectures and hyperparameters and transfer knowledge from small datasets to large datasets.
	
	\item Extensive experiments on real-world datasets demonstrate that our proposed approach can efficiently search configurations in designed space. 
	Furthermore,  results of ablation and case study show the superiority of our method. 
\end{itemize}

\section{Related Work}
\label{sec::related}

\subsection{Automated Machine Learning (AutoML)}
\label{sec::relatedwork1}
Automated Machine Learning  (AutoML)~\cite{hutter2019automatedck,quanming2018auto} refers to a type of method that can learn models adaptively to various tasks. 
Recently, AutoML has achieved great success in designing the state-of-the-art model for various applications such as image classification and segmentation~\cite{zoph2017neural,tan2019efficientnet,liu2019auto}, natural language modeling~\cite{so2019evolved},  and knowledge graph embedding~\cite{zhang2019autokge}.

AutoML can be used in mainly two fields: \textit{Neural Architecture Search} (NAS) and \textit{Hyperparameter Optimization} (HPO). 

\begin{itemize}[leftmargin=*]
\item 
NAS~\cite{zoph2017neural,pham2018efficient,li2019random} splits architectures into several components and searches for each part of the architecture to achieve the whole part.
DARTS~\cite{liu2018darts} uses gradient descent on continuous relaxation of the architecture representation, while NASP~\cite{yao2019differentiable} improves DARTS by including proximal gradient descent on architectures. 

\item 
HPO~\cite{bergstra2012random,li2017hyperband,feurer2019auto}, usually tuning the hyperparameters of a given architecture, always plays an important role in finding the best hyperparameters for the task.
Random Search is a frequently used method in HPO for finding proper hyperparameters. 
Algorithms for HP search based on model have been developed for acceleration~\cite{claesen2015hyperparameter}, including Bayesian Optimization (BO) methods like Hyperopt~\cite{bergstra2013hyperopt},BOHB~\cite{hutter2011sequential} and BORE~\cite{tiao2021bore}, etc.
\end{itemize}

Recent studies on AutoML have shown that incorporating HPO into the NAS process can lead to better performance and a more effective exploration of the search space. 
For example, a study on ResNet~\cite{zela2018towards} showed that considering the NAS process as HPO can lead to improved results. 
Other works, such as AutoHAS~\cite{dong2020autohas}, have explored the idea of considering hyperparameters as a choice in the architecture space. 
FEATHERS~\cite{seng2023feathers} has focused on the joint search problem in Federated Learning. 
ST-NAS~\cite{cai2021stnas} uses weight-sharing NAS on architecture space and consider HP as part of architecture encoding.
These studies demonstrate the potential of joint search on hyperparameters and architectures in improving the performance and efficiency of machine learning models.

\subsection{Collaborative Filtering (CF)}
\label{sec::relatedwork2}
\subsubsection{Classical CF Models}

Collaborative Filtering (CF)~\cite{sarwar2001item} is the most fundamental solution for Recommender Systems (RecSys). 
CF models are usually designed to learn user preferences based on the history of user-item interaction. 
Matrix Factorization (MF)~\cite{koren2008factorization} generates IDs of users and items, using a high-dimensional vector to represent the location of users and items' features. 
The inner product is used as interaction function, calculating the similarity of user/item vectors. 
%Besides, the interaction history can also be encoded to represent users and items by projecting them into high-dimensional space. 
The MF-based method has been demonstrated effective in SVD++~\cite{koren2008factorization} and FISM~\cite{kabbur2013fism}.
NCF~\cite{he2017neural} applied neural networks to building CF models, using a fused model with MF and multi-layer perceptron (MLP) as an interaction function, taking user/item embeddings as input, and inferring preference scores. 
JNCF~\cite{chen2019joint} extended NCF by using user/item history to replace user/item ID as the input encoding.

Recently, the user-item interaction matrix can also be considered as a bipartite graph, thus Graph Neural Networks (GNNs)~\cite{gao2023survey} are also applied to solve CF tasks for their ability to capture the high-order relationship between users and items~\cite{wang2019neural}.
They consider both users and items as nodes and the interaction of users and items as edges in bipartite graph. 
For example, PinSage\cite{ying2018graph} uses sampling on a graph according to node (user/item) degrees, and learn the parameters of GNN on these sampled graphs.
NGCF~\cite{wang2019neural} uses a message-passing function on both users themselves and their neighbor items and collects both the information to build proper user and item embeddings. 
LightGCN~\cite{he2020lightgcn} generates embeddings for users and items with simple SGC layers.

\subsubsection{AutoML for CF}
Recently, AutoML has been frequently used in CF tasks, 
aiming at finding proper hyperparameters and architectures for different tasks~\cite{hutter2019automatedck,quanming2018auto}.
Hyperparameter Optimization (HPO) has been applied on the embedding dimension of RecSys models.
%These works focus on features of users and items, and the components in their search space include embedding dimension and interaction function.
For example, 
AutoDim~\cite{zhao2021autodim} searches embedding dimension in different fields, aiming at assigning embedding dimension for duplicated content.
PEP~\cite{liu2021learnable} used learnable thresholds to identify the importance of parameters in the embedding matrix,
and trains the embedding matrix and thresholds by sub-gradient decent. 
AutoFIS~\cite{liu2020autofis} is designed to learn feature interactions by adding an attention gate to  every potential feature interaction.
Most of these works tune the embedding dimension adaptively on Recommender Systems tasks, mostly in Click-Through-Rate (CTR) tasks.

Neural Architecture Search (NAS)  has also been applied on CF tasks, including SIF~\cite{yao2020searching} AutoCF~\cite{gao2021efficient} and~\cite{wang2022profiling}.
In detail, 
SIF adopts the one-shot architecture search for adaptive interaction function in the CF model.
AutoCF designs an architecture space of neural network CF models, and the search space is divided into four parts: encoding function, embedding function, interaction function and prediction function. 
AutoCF selects an architecture and its hyperparameters in the space with a performance predictor. 
Hyperparameters are considered as a discrete search component in search space, neglecting its continuous characteristic. 
\cite{wang2022profiling} designs the search space on graph-based models,  which also uses random search on the reduced search space.

\section{Search Problem}
\label{sec::problem}
As mentioned in the introduction, 
finding a proper CF model should be considered as a joint search problem on hyperparameters and architectures.
We propose to use AutoML to find architectures and their proper hyperparameters efficiently. 
The joint search problem on CF hyperparameters and architectures can be modeled as a bilevel optimization problem as follows:

\begin{mdefinition}[Joint Automated Hyperparameters and Architecture Search for CF] 
Let $f^{*}$ denote the proper CF model, and then the joint search problem for CF can be formulated as:
\label{prob::def}
\begin{align}
    \alpha^*, h^* 
    & = \max_{\alpha \in \mathcal{A}, h \in \mathcal{H}}
    \mathcal{M}(f(\mathbf{P}^{*}; \alpha, h), \mathcal{S}_{\textit{val}}),
    \label{eq:autocf1}
    \\
    \text{\;s.t.\;} 
    & \mathbf{P}^* 
    = \argmax_\mathbf{P} \mathcal{M}(f(\mathbf{P}; \alpha, h), \mathcal{S}_{\textit{tra}}) \label{eq:autocf2} .
\end{align}
where $\mathcal{H}$ contains all possible choices of hyperparameters $h$, 
where $\mathcal{A}$ contains all possible choices of architectures $\alpha$, 
$\mathcal{S}_{\textit{val}}$ and $\mathcal{S}_{\textit{tra}}$ denote the training and validation datasets, 
$\mathbf{P}$ denotes the learnable parameters of the CF architecture $\alpha$, 
and $\mathcal{M}$ denotes the performance measurement, such as \text{Recall} and \text{NDCG}.
\end{mdefinition}

We encounter the following key challenges in effectively and efficiently solving the search problem:
Firstly, 
the joint search space must contains a wide range of architectural operations within $\mathcal{A}$, as well as frequently utilized hyperparameters in the learning stage within $\mathcal{H}$. 
Since this space contains various types of components, including continuous hyperparameters and categorical architectures, it is essential to appropriately encode the joint search space for effective exploration.
Secondly, considering the dependency between hyperparameters and architectures, the search strategy should be robust and efficient, meeting the accuracy and efficiency requirements of real-world applications. 
Compared to previous AutoML works on CF tasks, our method is the first to consider a joint search on hyperparameters and architectures on CF tasks.

\subsection{Architecture Space: $\mathcal{A}$}
In this paper, the general architecture of CF models can be separated into four parts~\cite{gao2021efficient}: 
Input Features, Feature Embedding, Interaction Function, and Prediction Function.
Based on the frequently used operations, we build the architecture space $\mathcal{A}$, illustrated in Table~\ref{tab::architecture_space}.

\begin{table}[t]
    \centering
    \caption{The operations we use for architecture space.}
    \label{tab::architecture_space}
%    \vspace{-10pt}
    \setlength\tabcolsep{2pt}
    \begin{tabular}{c|c|C{90px}}
        \toprule
        \bf  Architecture  & \multicolumn{2}{c}{\bf Operations} \\ \midrule
        \multirow{2}{*}{\bf Input Features}  &  User &\texttt{ID}, \texttt{H}  \\ \cmidrule{2-3}
        & Item &\texttt{ID},\, \texttt{H}  \\ \midrule
        {\bf Features Embedding}   &  NN-based  &\texttt{Mat},\, \texttt{MLP} \\ \cmidrule{2-3}
        & Graph-based  & \texttt{GraphSAGE}, \,\texttt{HadamardGCN} ,\, \texttt{SGC}  \\ 
        \midrule
        \bf  Interaction Function   & \multicolumn{2}{c}{ \texttt{multiply},\, \texttt{minus},\, \texttt{min},\, \texttt{max},\, \texttt{concat}}   \\ 
        \midrule
        \bf Prediction Function &\multicolumn{2}{c}{ \texttt{SUM},\, \texttt{VEC}, \, \texttt{MLP} }\\ 
        \bottomrule
    \end{tabular}
\end{table}

\para{Input Features} 
The input features of CF architectures come from original data: user-item rating matrix. 
We can apply the interactions between users and items and map them to high-dimension vectors. 
There are two manners for encoding users and items as input features: one-hot encoding (\texttt{ID}) and multi-hot encoding (\texttt{H}). 
As for one-hot encoding (\texttt{ID}), we consider the reorganized id of both users and items as input. Since the number of users and items is different, we should maintain two matrices when we generate these features.
As for multi-hot encoding (\texttt{H}), we can consider the interaction of users and items. 
A user vector can be encoded in several places by its historical interactions with different items. 
The items can be encoded in the same way.

\begin{figure*}[t]
	\centering
	\begin{subfigure}[t]{0.29\textwidth}
		\centering
		\includegraphics[width=\linewidth]{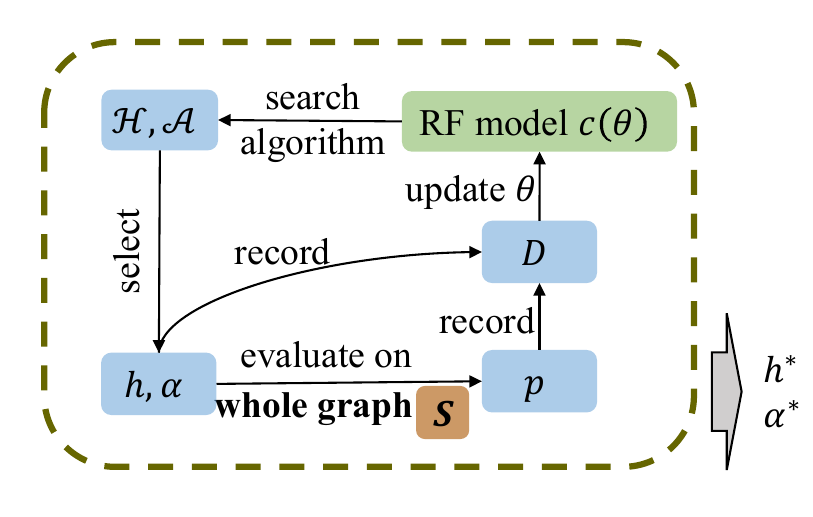}
%		\vspace{-13pt}
		\caption{Conventional methods}
		\label{fig:search_one_stage}
	\end{subfigure}
	\begin{subfigure}[t]{0.69\textwidth}
		\centering
		\includegraphics[width=\linewidth]{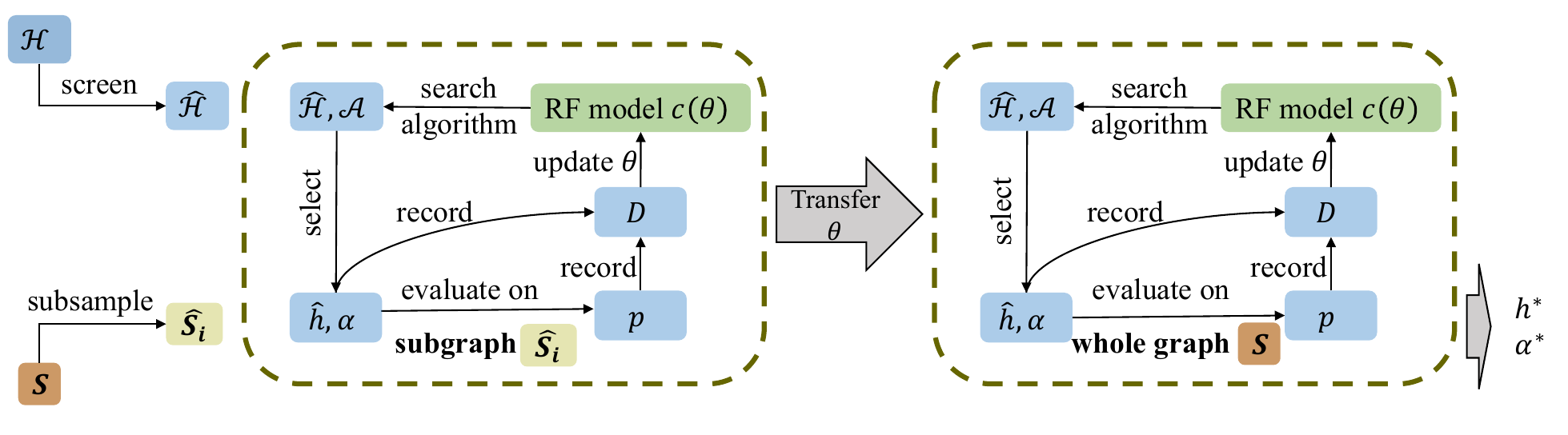}
%		\vspace{-10pt}
		\caption{Our method (Algorithm~\ref{alg::search_alg})}
		\label{fig:search_two_stage}
	\end{subfigure}
%	\vspace{-6pt}
	\caption{Joint search strategy of conventional methods and our method.}
	\label{fig:search_strategy_whole}
\end{figure*}

\begin{table*}[t]
	\centering
	\caption{Origin hyperparameter space, discretized values, and shrunk range after using screening method in Section~\ref{sec::strategy::screen_hp}. }
	\label{tab:hp_space_origin}
%	\vspace{-10pt}
	\begin{tabular}{c|c|c|c}
		\toprule
		Hyperparameter & Original range & Discrete values & Shrunk range \\
		\midrule
		optimizer 		& Adagrad, Adam, SGD & Adagrad, Adam, SGD & Adagrad, Adam \\ 
		learning rate   &  [1e-6, 1e-1]   &  $\{ 10^{-6}, 10^{-5},\dots,10^{0}\}$ & [1e-5, 1e-2]       \\ 
		embedding dimension   & [1,512] &    $\{2^{0},2^{1},\dots, 2^{9}\}$   &    [2,64]    \\ 
		weight decay 	& [1e-5, 1e-1] 	 & $\{ 10^{-5}, 10^{-4},\dots,10^{-1}\}$ &  1e-1      \\ 
		batch size 	& [500, 5000] 	 & $\{ 500, 1000,\dots, 5000 \}$    & $\{ 2000 \}$    \\ 
		\bottomrule
	\end{tabular}
\end{table*}

\para{Feature Embedding}
The function of feature embedding maps the input encoding with high dimension into vectors with lower dimension.
According to designs in Section~\ref{sec::relatedwork2}, we can elaborate the embedding manners in two categories: Neural Networks based (NN-based) embeddings and Graph Neural Networks based (Graph-based) embeddings. The embedding function is related to the size of input features. 
As for NN-based methods, a frequently used method of calculation is \texttt{Mat}, mainly consists of a \textit{lookup-table} in \texttt{ID} level, and mean pooling on both users/items sides. 
We can also use a multi-layer perceptron (\texttt{MLP}) for each user and item side, helping convert multi-hot interactions into low-dimensional vectors.
As for Graph-based methods, 
the recent advances in GNNs use more complex graph neural networks to aggregate the neighborhoods, such as \texttt{GraphSAGE}~\cite{hamilton2017inductive}, \texttt{HadamardGNN}~\cite{wang2019neural}, and \texttt{SGC}~\cite{he2020lightgcn} etc. 
%These methods also take interaction history as input (history can be regarded as the neighboring relation). 

\para{Interaction Function}
The interaction function calculates the relevance between a given user and an item. 
In this operation, the output is a vector affected by the embeddings of the user and item. The inner product is frequently used in many research works on CF tasks. We split the inner product and consider an element-wise product as an essential operation, which is noted as \texttt{multiply}.
In coding level, we can also use \texttt{minus}, \texttt{max}, \texttt{min} and \texttt{concat}. 
They help us join users and items with different element-wise calculations.

\para{Prediction Function}
This operation stage helps turn the output of the interaction function into an inference of similarity. 
As for the output vector of a specific interaction, a simple way is to use summation on the output vector. 
Thus, \texttt{multiply+SUM} can be considered the inner product in this way. 
Besides, we use a weight vector with learnable parameters, noted as \texttt{VEC}. 
Multi-layer perceptron (\texttt{MLP}) can also be used for more complex prediction on similarity.

\subsection{Hyperparameter Space: $\mathcal{H}$}
Besides the model architecture, 
the hyperparameter (HP) setting also plays an essential role in determining the performance of the CF model.
The used components for hyperparameter space are illustrated in the first column of Table~\ref{tab:hp_space_origin}.

The CF model, like any machine learning model, consists of standard hyperparameters such as learning rate and batch size. 
Specifically, excessively high learning rates can hinder convergence, while overly low values result in slow optimization. 
%Similarly, the batch size represents a trade-off between efficiency and effectiveness, where selecting a larger batch size may improve computational efficiency but could potentially sacrifice model performance.
The choice of batch size is also a trade-off between efficiency and effectiveness.

In addition, CF model has specific and essential hyperparameters, which may not be so sensitive in other machine learning models. 
Embedding dimension for users and items influence the representative ability of CF models.
Besides, the embedding size determines the model's capacity to store all information of users and items. 
In general, too-large embedding dimension leads to over-fitting, and too-small embedding dimension cannot fit the complex user-item interaction data.
The regularization term is always adopted to address the over-fitting problem.

\begin{figure*}[ht]
	\centering
	\begin{subfigure}[t]{0.29\textwidth} 
		\centering
		\includegraphics[width=\textwidth]{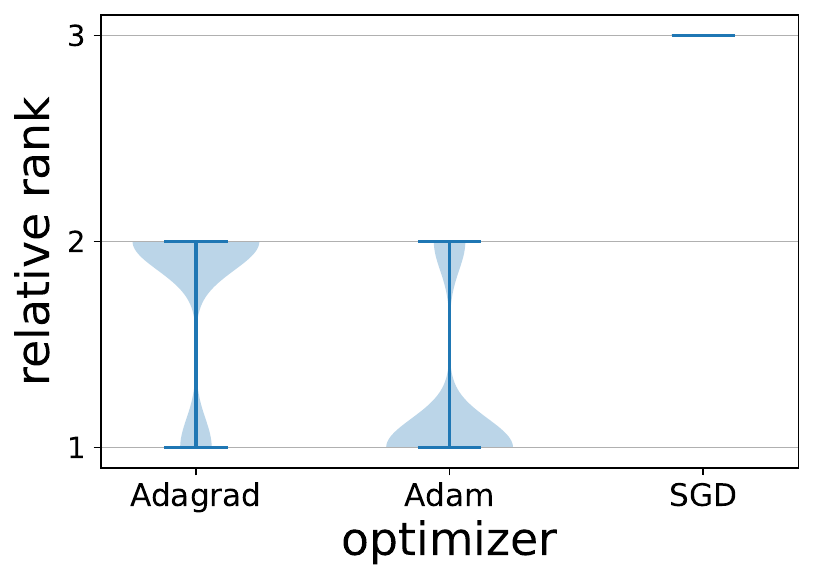}
%		\vspace{-15pt}
	\end{subfigure}
	\begin{subfigure}[t]{0.29\textwidth}
		\centering
		\includegraphics[width=\textwidth]{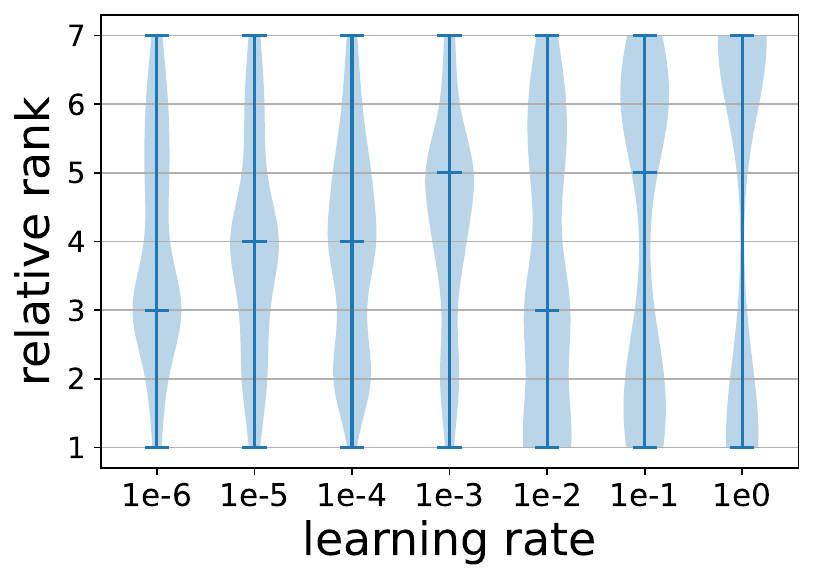}
%		\vspace{-15pt}
	\end{subfigure}
	\begin{subfigure}[t]{0.29\textwidth} 
		\centering
		\includegraphics[width=\textwidth]{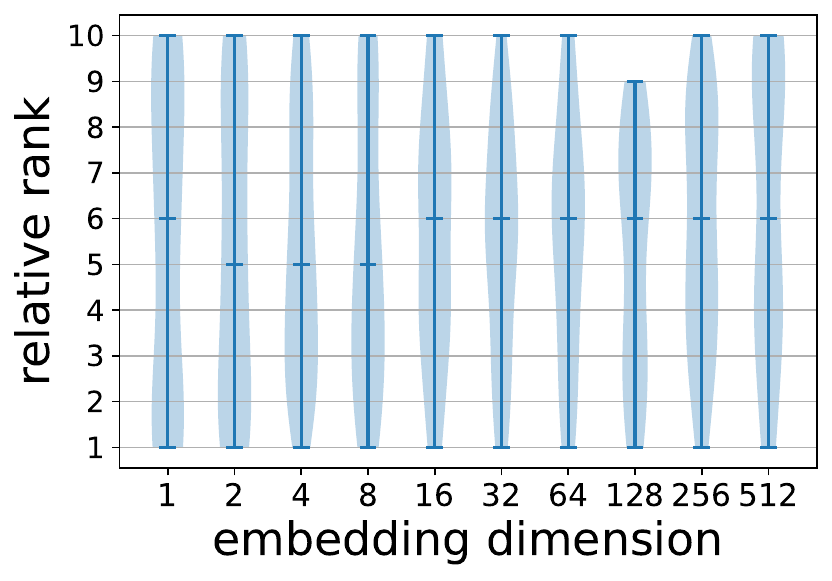}
%		\vspace{-15pt}
	\end{subfigure}
	\newline
	\begin{subfigure}[t]{0.29\textwidth} 
		\centering
		\includegraphics[width=\textwidth]{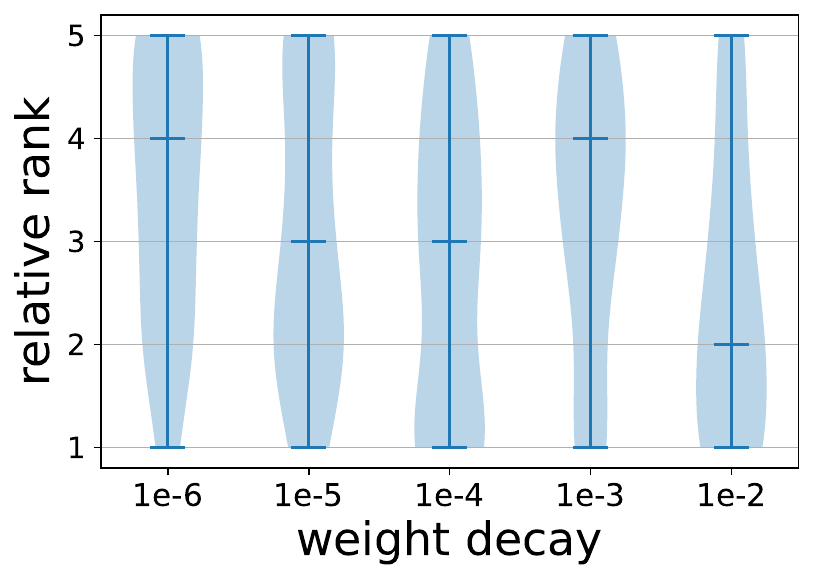}
%		\vspace{-15pt}
	\end{subfigure}
	\begin{subfigure}[t]{0.29\textwidth} 
		\centering
		\includegraphics[width=\textwidth]{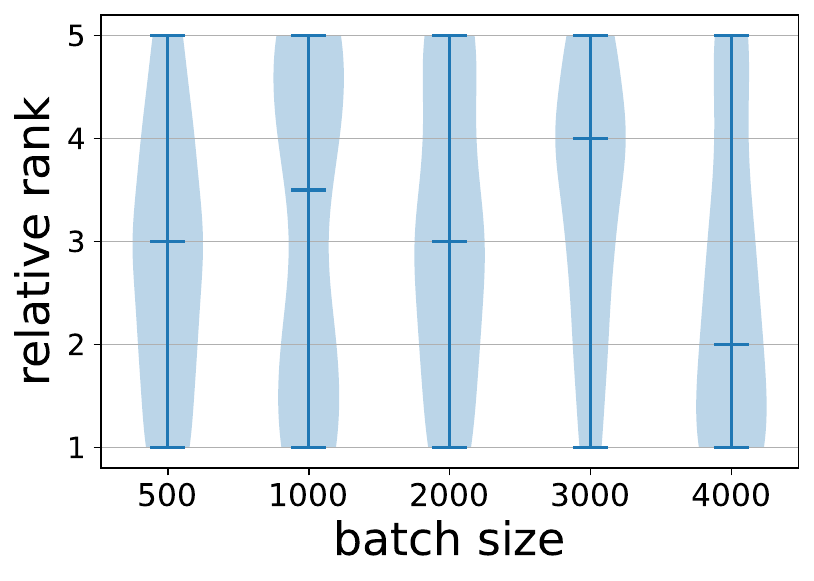}
%		\vspace{-15pt}
	\end{subfigure}
%	\vspace{-10pt}
	\caption{Ranking distribution of hyperparameters.}
	\label{fig:perf_rank_hp_space}
\end{figure*}

\section{Search Strategy}
As mentioned in Section~\ref{sec::intro}, 
joint search on hyperparameters and architectures have two challenges in designing a search strategy effectively and efficiently: the large search space and the high evaluation costs on large network datasets. 
Previous research works on model design with joint search space of hyperparameters and architectures such as~\cite{dong2020autohas, zela2018towards} consider the search problem in the manner of  Figure~\ref{fig:search_one_stage}. 
They considered the search component in space jointly and used different search algorithms to make the proper choice. 
The search procedure can be costly on a large search space and dataset. 
Therefore, addressing the challenges of a vast search space and the costly evaluation process is crucial.

The first challenge means the joint search space of $\mathcal{H}$ and $\mathcal{A}$ described in Section~\ref{sec::problem} is large for accurate search. 
Thus, we need to reduce the search space. In practice, we choose to screen $\mathcal{H}$ choices by comparing relative ranking in controlled variable experiments, which is explained in Section~\ref{sec::strategy::screen_hp}.
The second challenge means the evaluation time cost in Equation~(\ref{eq:autocf2}) is high. 
A significant validation time will lower the search efficiency. 
Thus, we apply a sampling method on datasets by interaction frequency, elaborated in Section~\ref{sec::strategy::arch_transfer}.

In comparison to conventional joint search problem in Figure~\ref{fig:search_one_stage}, 
we design a two-stage search algorithm in Figure~\ref{fig:search_two_stage}. 
We use Random Forest (RF) Regressor, a surrogate model, to improve the efficiency of the search algorithm. 
To transfer the knowledge, including the relation modeling between hyperparameters, architectures, and evaluated performance, we learn the surrogate model's parameters $\theta$ in the first stage, and we use $\theta$ as initialization of RF model in the second stage. 
Our search algorithm is shown in Algorithm~\ref{alg::search_alg} and described in detail in Section~\ref{sec::strategy::arch_search}.
Besies, we have a discussion on our choices in Section~\ref{sec::strategy::discussion}, elaborating how we solve the challenge in the joint search problem.

\subsection{Screening Hyperparameter Choices}
\label{sec::strategy::screen_hp}

We screen the hyperparameter (HP) choices from $\mathcal{H}$ to $\hat{\mathcal{H}}$ with two techniques.
First, we shrink the HP space by comparing relative performance ranking of a special HP while fixing the others. 
After we get the ranking distribution of different HPs, we find the performance distribution among different choices of a HP, thus we may find the proper range or shrunk set of a given HP.
Second, we decouple the HP space by calculating the consistency of different HP. 
If the consistency of a HP is high, that means the performance can change positively or negatively by only alternating this HP. 
Thus, this HP can be tuning separately neglecting its relation with other HPs in HP set.

\subsubsection{Shrink the hyperparameter space}
The screening method on hyperparameter space is based on analysis of ranking distribution on the performance with fixed value for a certain HP and random choices for other HPs and architectures.
In this part, we denote a selection of hyperparameter $h \in \mathcal{H}$ as a vector,  noted as $h = \left(h^{(1)}, h^{(2)},\dots, h^{(n)}\right)$.
%The range for $i$-th HP $h^{(n)}$ is $\mathcal{H}_i$. 
For instance, $h^{(1)}$ means optimizer, and $h^{(2)}$ means learning rate.

To obtain the ranking distribution of a certain HP $h^{(i)}$, we start with a controlled variable experiment.
We vary $h^{(i)}$ in discrete values as the third column in Table~\ref{tab:hp_space_origin}, and we vary other HPs in original range as the second column.
Specifically, given $H_i$ as a discrete set of HP $h^{(i)}$, we choose a value $\lambda \in H_i$ and we can obtain the ranking $\texttt{rank}(\mathsf{h}, \lambda)$ of the anchor HP $ \mathsf{h} \in \mathcal{H}_i $ by fixing other HPs except the $i$-th HP. 

To ensure a fair evaluation of different architecture, 
we traverse the architecture space and calculate rank of performance with different configurations, then we can get the distribution of a type of HP. 
The relative performance ranking with different HPs is shown as violin plots in Figure~\ref{fig:perf_rank_hp_space}.
In this figure, we can get $\mathcal{\hat H}$ through the distribution of different HP values. 
We learn that the proper choice for optimizer can be shrunk to Adam and Adagrad; Proper range for learning rate is (1e-5, 1e-2); Proper range for embedding dimension can be reduced to [2, 64]; And we can fix weight decay in our experiments. 
We demonstrate the conclusion in the fourth column in Table~\ref{tab:hp_space_origin}.

\subsubsection{Decouple the hyperparameter space}
\begin{figure}[t]
	\centering
	\includegraphics[width=0.4\textwidth]{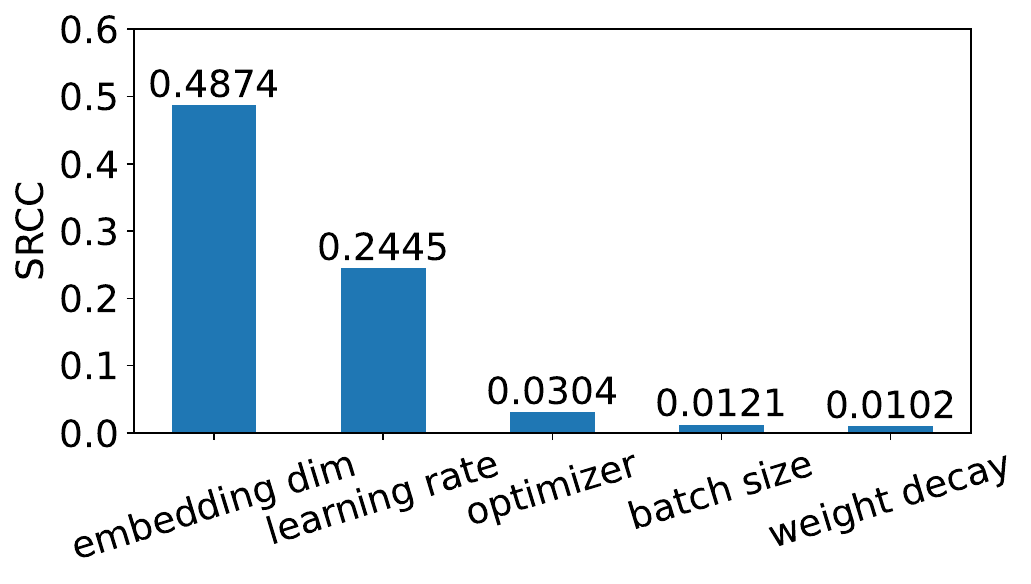}
%	\vspace{-10pt}
	\caption{Consistency of each hyperparameters.}
	\label{fig::srcc_hp_decouple}
\end{figure}

To decouple the search space, 
we consider the consistency of the ranking of hyperparameters when only alternating a given hyperparameter.
For the $i$-th element $h^{(i)}$ of $h\in \mathcal{H}$, 
we can change different values for $h^{(i)}$, 
and then we can decouple the search procedure of the $i$-th hyperparameter with others.
We use Spearman Rank-order Correlation Coefficient (\texttt{SRCC}) to show the consistency of various types of HPs, which is definited in Equation~(\ref{eq:srcc_def}).

\begin{equation}\label{eq:srcc_def}
	\texttt{SRCC}(\lambda_1, \lambda_2) = 1-\frac{\sum_{h\in\mathcal{H}_i}|\texttt{rank}(h,\lambda_1)-\texttt{rank}(h,\lambda_2)|^2}{|\mathcal{H}_i|\cdot(|\mathcal{H}_i|^2-1)}.
\end{equation}
where $|\mathcal{H}_i|$ means the number of anchor hyperparameters in $\mathcal{H}_i$. 
\texttt{SRCC} demonstrates the matching rate of rankings for anchor hyperparameters in $\mathcal{H}_i$ with respect to $h^{(i)} = \lambda_2$ and $h^{(i)} = \lambda_2$.

The \texttt{SRCC} of the $i$-th HP is evaluated by the average of $\texttt{SRCC}(\lambda_1, \lambda_2)$ among different pairs of $\lambda \in H_i$, as is shown in Equation~(\ref{eq:srcc_avg}).

\begin{equation}\label{eq:srcc_avg}
	\texttt{SRCC}_{i} = \frac{1}{|H_i|^2} \sum_{(\lambda_1, \lambda_2)\in H_i\times H_i}\texttt{SRCC}(\lambda_1, \lambda_2).
\end{equation}

%For the definition of \texttt{SRCC}, it will be in the range of $[-1,1]$, larger consistency indicates that alternating the value of the $i$-th HP (i.e. from $\theta_1$ to $\theta_2$) does not influence a lot on the ranking of other hyperparameters.

The \texttt{SRCC} results is demonstrated in Figure~\ref{fig::srcc_hp_decouple}, 
we can directly find that the important hyperparameter with higher \texttt{SRCC} has a more linear relationship with its values. 
Since high embedding dimension model is time-costly during training and evaluation, we decide to use lower dimension in the first stage to reduce validation time, and then raise them on the original dataset to get better performance. 

To summarize, we shrink the range of HP search space and find the consistency of different HPs. 
The shrunk space shown in Table~\ref{tab:hp_space_origin} help us search more accurately, and the consistency analysis on performance ranking help us find the dependency between different HPs, thus we can tune HP with high consistency separately.

\subsection{Evaluating Architecture with Sampling} 
\label{sec::strategy::arch_transfer}

To evaluate architecture more efficiently,  we collect performance information evaluated from subgraphs,  since the subgraph can approximate the properties of whole graph~\cite{zhang2022kgtuner}.
In this part,  we introduce our frequency-based sampling method, and then we show the transfer ability of subgraphs by testing the consistency of architectures' performance from subsampled dataset to origin dataset.

\subsubsection{Matrix sampling method}
Since dataset for CF based on interaction of records, our sampling method is based on items' appearance frequency.
That is, we can subsample the original dataset when we preserve part of the bipartite graph, 
and the relative performance on smaller datasets should have a similar consistency (i.e. ranking distribution of performance on $\mathcal{S}_{\textit{val}}$ and $\hat{\mathcal{S}}_{\textit{val}}$).

The matrix subsample algorithm is demonstrated in Algorithm~\ref{alg::subsample}.
First, we set the number of user-item interactions to be preserved first,  which can be controlled by a sample ratio $\gamma$, $\gamma\in (0,1)$.
We calculate the interactions for each item, and then we preserve the item with a higher frequency.
The items can be chosen in a fixed list (i.e. \texttt{topk}, Line 6-7 in Algorithm~\ref{alg::subsample}), 
or the interaction frequency count of items can be normalized to a probability, 
then different items have corresponding possibility to be preserved (i.e. \texttt{distribute}, Line 8-10 in Algorithm~\ref{alg::subsample}).

\subsubsection{Transfer ability of architecture on subsampling matrix}

To ensure that the relative performance ranking on subsampled datasets is similar to that on original datasets, we need to test the consistency of architecture ranking on different datasets.
We evaluate the transfer ability among from subgraph to whole graph by \texttt{SRCC}.

For a given value of $\gamma$, we choose to select a sample set of architecture from $A_{\gamma}\in\mathcal{A}$.
Then we evaluate them on a subsampled dataset $\hat{\mathcal{S}}$ and origin dataset $\mathcal{S}$. The relative rank of $\alpha\in A_{\gamma}$ on $\hat{\mathcal{S}}$ and $\mathcal{S}$ is noted as $\texttt{rank}(\alpha, \hat{\mathcal{S}})$ and $\texttt{rank}(\alpha, \mathcal{S})$.
%We first get the rank of a fixed architecture on both smaller datasets as a group of pair data, and then we calculate \texttt{SRCC}. 

\begin{equation}
	\texttt{SRCC}_{\gamma} = 1-\frac{\sum_{\alpha\in A_{\gamma}}|\texttt{rank}(\alpha, \hat{\mathcal{S}})-\texttt{rank}(\alpha, \mathcal{S})|^2}{|A_{\gamma}|\cdot(|A_{\gamma}|^2-1)}.
\end{equation}

\begin{figure}[t]
	\centering
	\includegraphics[width=0.42\textwidth]{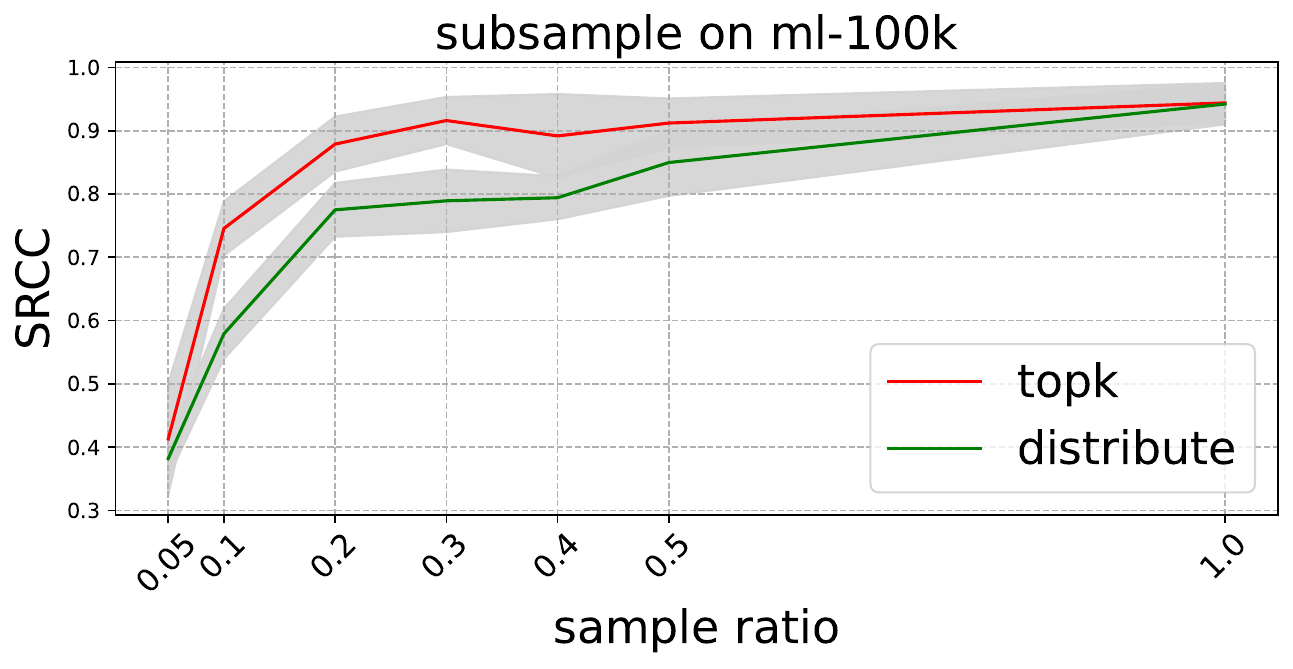}
%	\vspace{-10pt}
	\caption{Consistency of different sample ratio. The light grey aera is the standard deviation of \texttt{SRCC} on different $\hat{\mathcal{S}}$. }
	\label{fig:srcc_sp_ml100k}
\end{figure}

We can choose different subsampled dataset $\hat{\mathcal{S}}$ to get average consistency.
As is demonstrated in Figure~\ref{fig:srcc_sp_ml100k}, 
sample ratio with higher \texttt{SRCC} has better transfer ability among graphs with sample mode \texttt{topk}, and the proper sample ratio should be in $\gamma \in [0.2,1)$.

To summarize, through the sampling method, the evaluation cost will be reduced, and thus the search efficiency is improved. 
Since the ranking distribution among subgraphs is similar to that of the original dataset, we can transfer the evaluation modeling from small to large dataset.

\begin{algorithm}[t]
	\caption{\texttt{MatrixSample} Sampling Algorithm}
	\label{alg::subsample}
	\begin{flushleft}
		\textbf{Input}: Matrix Dataset $\mathcal{S}\in\mathbb{R}^{M \times N}$ ($M$: number of users; $N$: number of items), sample ratio $\gamma\in(0,1)$, subsample mode (\texttt{topk} or \texttt{distribute}) \\
		\textbf{Output}: Subsampled Matrix Dataset $\hat{\mathcal{S}}$ \\
	\end{flushleft}
	\begin{algorithmic}[1] %
		\STATE Initialize item frequency set $f\leftarrow \{\}$ ;
		\FOR{ $j = 1,2,\cdots, N$}
		\STATE Calculate the users interacted with item $j$, record it as $f_j$ ;
		\STATE $f \leftarrow f \bigcup f_j $;
		\ENDFOR
		\IF{subsample mode == \texttt{topk}}
		\STATE Rank $f$, choose top $\gamma N$ items in $\hat{f}$, neglect  unrelated users;
		\ELSIF{subsample mode == \texttt{distribute}}
		\STATE Calculate probability $\beta_j=f_j/\sum_{k=1}^{N}f_k$;
		\STATE Select $\hat{f}$ with respect to $\{\beta_j\}$;
		\ENDIF
		\STATE Construct $\hat{\mathcal{S}}$ with $f$;
		\STATE \textbf{return} $\hat{\mathcal{S}}$.
	\end{algorithmic}
\end{algorithm}

\subsection{Two-stage Joint Hyperparameter and Architecture Search} 
\label{sec::strategy::arch_search}

%\subsubsection{Efficient search algorithm}
As discussed above, 
the evaluation cost can be highly reduced with sampling methods. 
Since the sampling method ensure the transfer ability from subgraphs to whole graphs, 
we propose a two-stage joint search algorithm, shown in  Algorithm~\ref{alg::search_alg}, and the framework is also shown in Figure~\ref{fig:search_two_stage}. 
The main notations in algorithm can be found in Table~\ref{tab::notations} in Appendix. 
We also compare our method with conventional method in Figure~\ref{fig:search_procedure}.

To briefly summarize our algorithm:
In the first stage (Lines 3-14), 
we sample several subgraphs with frequency-based methods, 
and preserve $\gamma=0.2$ of interactions from the original rating matrix.
Based our understanding of hyperparameters in Section~\ref{sec::strategy::screen_hp}, 
we select architecture and its hyperparameters in our reduced hyperparameter space.

We use a surrogate model to find proper architecture and hyperparameters to improve search efficiency, noted as $c(\cdot, \theta)$, $\theta$ is the parameter of the model.
After we get evaluations of architecture and hyperparameters, we update parameters $\theta$ of $c$.
In our framework, we choose BORE~\cite{tiao2021bore} as a surrogate and Random Forest (RF) as the regressor to modulate the relation between search components and performance. 
Details of the BORE+RF search algorithm (\texttt{RFSurrogate}) is shown in Algorithm~\ref{alg::surrogate} in Appendix.

We first transfer the parameters of $c$ trained in the first stage by collected configurations and performance.
Then, in the second stage (Lines 15-22), 
we select architectures and hyperparameters and evaluate them in the original dataset.
Besides, we increase the embedding dimension to reach better performance.
Similarly, the configurations and performance are recorded for the surrogate model (RF+BORE), 
which can give us the next proper configuration.
Finally, after two-stage learning on the subsampled and original datasets, we get the final output of architecture and hyperparameters as the best choice of architecture and its hyperparameters.

\begin{algorithm}[t]
	\caption{Joint Search Algorithm.}
	\begin{flushleft}
	\label{alg::search_alg}
	\textbf{Input}: Train/valid set $S_{\textit{tra}}, S_{\textit{val}}$, 
	hyperparameter space $\mathcal{H} $, architecture space $\mathcal{A}$, 
	sample number $L$, surrogate model $c(\cdot, \theta)$ \\
	\textbf{Output}: Best hyperparameters $h^*$ , architecture $\alpha^*$\\
	\end{flushleft}

	\begin{algorithmic}[1] %
		\STATE Screen $\mathcal{H} $ to $\hat{\mathcal{H}}$   according to Section~\ref{sec::strategy::screen_hp};
		\STATE Initialize configuration set $D \leftarrow \{ \},k,j\leftarrow 0 $ ;\\
		\%\textbf{stage one} start
		\FOR{ $i = 1,2,\cdots, L$ }
		\STATE Sample matrix  $\hat{\mathcal{S}}_i=\texttt{MatrixSample}(\mathcal{S},\gamma)$;
		\STATE Split train/valid/test dataset $\hat{\mathcal{S}}_i=\hat{\mathcal{S}}_{i,\textit{tra}}\bigcup\hat{\mathcal{S}}_{i,\textit{val}}\bigcup\hat{\mathcal{S}}_{i, \textit{tst}}$;
		\STATE Initialize surrogate model $c(\cdot;\theta)$;
		\WHILE{not converge}
		\STATE Select $\alpha_k$ and $h_k$  by \texttt{RFSurrogate};
		\STATE Evaluate $p_k \leftarrow \mathcal{M}(f(\textbf{P*};\alpha_k, h_k), \hat{\mathcal{S}}_{i,\textit{val}})$ ;
		\STATE Save to set $D \leftarrow D \bigcup \{ \alpha_k, h_k, p_k \} $;
		\STATE Update $c(\cdot;\theta)$ with $D$ by \texttt{RFSurrogate};
		\STATE $k\leftarrow k+1$;
		\ENDWHILE
		\ENDFOR
		
		\STATE Transfer parameters $\theta$ in $c$;

		\%\textbf{stage two} start 
		\WHILE{not converge}
		\STATE Select $\alpha_j$ and $h_j$ by \texttt{RFSurrogate}; 
		\STATE Evaluate $p_j \leftarrow \mathcal{M}(f(\textbf{P*};\alpha_j, h_j), \mathcal{S}_{\textit{val}})$ ;
		\STATE Save to set $D \leftarrow D \bigcup \{ \alpha_j, h_j, p_j \} $;
		\STATE Update $c(\cdot;\theta)$ with $D$ by \texttt{RFSurrogate};
		\STATE $j\leftarrow j+1$;
		\ENDWHILE
		\STATE \textbf{return} $\alpha^*, h^*$.
	\end{algorithmic}
\end{algorithm}

\section{Experiments}
\label{sec::exp}

Extensive experiments are performed to evaluate the performance of our search strategy by answering the following several research questions:

\begin{itemize}[leftmargin=*]
	\item \textbf{RQ1}: How does our algorithm work in comparison to other CF models and automated model design works?
	\item \textbf{RQ2}: How efficiently does our search algorithm work in comparison to other typical search algorithms? 
	\item \textbf{RQ3}: What is the impact of every part of our design? 
	\item \textbf{RQ4}: How specific is our model for different CF tasks? 	
\end{itemize}

\begin{table*}[t]
%	\small
	\caption{Comparison of different methods on CF tasks.}
	\label{tab::model_perform}
%	\vspace{-10pt}
	\setlength\tabcolsep{2pt}
	\centering
	\begin{tabular}{cccccccccc}
		\toprule
		\multicolumn{2}{c}{\bf Dataset}   & \multicolumn{2}{c}{\bf MovieLens-100K} & \multicolumn{2}{c}{\bf MovieLens-1M} & \multicolumn{2}{c}{\bf Yelp} & \multicolumn{2}{c}{\bf Amazon-Book} \\ \midrule
		\multicolumn{2}{c}{\bf Metric} &\bf Recall@20 &  \bf NDCG@20 &\bf  Recall@20 &   \bf NDCG@20 &\bf Recall@20 &  \bf NDCG@20  &\bf  Recall@20 &   \bf NDCG@20  \\ \midrule
		&  MF~\cite{koren2009matrix}       	& 0.1145    &  0.1179	&  0.0896   &  0.0584	&  0.0307   &  0.0286 	&  0.0291   &   0.0213   \\
		&  FISM~\cite{kabbur2013fism}      	&  0.1434 	&  0.1422 	&  0.0995   &  0.0621 	&  0.0528   &  0.0316  	&  0.0302   &   0.0211   \\
		&  NCF~\cite{he2017neural}       	&  0.1980 	&  0.1521   &  0.1204   &  0.0684   &  0.0534   &  0.0335  	&  0.0315   &   0.0223   \\
		&  J-NCF~\cite{zheng2017joint}   	&  0.2016  	&  0.1448 	&  0.1265   &  0.0629 	&  0.0636   &  0.0393 	&  0.0351   &   0.0252   \\ \midrule
		&  Pinsage~\cite{ying2018graph}    	&  0.1561  	&  0.1421 	&  0.1354  	&  0.0631 	&  0.0561   &  0.0417 	&  0.0321   &   0.0239   \\
		&  NGCF~\cite{wang2019neural}       &  0.1654  	&  0.1479	&  0.1678   &  0.0852  	&  0.0573   &  0.0454  	&  0.0349   &   0.0247   \\
		&  LightGCN~\cite{he2020lightgcn}  	&  0.2342	&  0.1755	&  0.1637   &  0.0842 	&  0.0689   &  0.0471  	&  0.0355   &   0.0273   \\ \midrule
		&  SIF~\cite{yao2020searching} 		&  0.1935   &  0.1532	&  0.1294   &  0.0695  	&   0.0581  &  0.0459  	&  0.0350   &   0.0249   \\
		%		&  Profiling~\cite{wang2022profiling}  &  0.2743   &  0.2019  	&  -   &  -  	&  -   &  - 	&  -   &     -       \\
		&  AutoCF~\cite{gao2021efficient}  	&  0.2259   &  0.1703 	&  0.1309   &  0.0739 	&  0.0643   &  0.0467  	&  0.0354   &   0.0265   \\ 
%		&   Ours (w/o tune HP)  		&  0.2542   &  0.1821  	&  -   &  -  	&  -   &  - 	&  -   &     -       \\
		&  \bf Ours  						&  0.2647   &  0.1913	&  0.1787   &  0.0945  	&  0.0721   &  0.0482  	&  0.0365   &   0.0281   \\ 
		\midrule
%		& \bf Improvement   				&  13.02\% 	&  12.33 \% & 9.16 \% 	& 12.23\% 	&   4.64\%	&	 2.33\%	&	3.11\%	&   2.93\%   \\ 
		& \bf Improvement   				&  13.02\% 	&  9.00 \% & 6.50 \% 	& 10.92\% 	&   4.64\%	&	 2.33\%	&	2.82\%	&   2.93\%   \\ 
		\bottomrule
	\end{tabular}
\end{table*}

\begin{table*}[ht]
%	\small
	\centering
	\caption{Top 3 architectures for each datasets.}
	\label{tab::case_study_top}
%	\vspace{-10pt}
	\begin{tabular}{c|ccc}
		\toprule
		\bf  Dataset & Top-1 & Top-2 &  Top-3  \\  \midrule
		\bf   ML-100K    &  \texttt{$\langle$H,H,SGC,SGC,min,VEC$\rangle$}   &      \texttt{$\langle$ID,ID,SGC,SGC,multiply,SUM$\rangle$}         &   \texttt{$\langle$H,H,Mat,Mat,multiply,VEC$\rangle$}    \\
		\bf     ML-1M     &  \texttt{$\langle$H,H,SGC,SGC,min,VEC$\rangle$}   &  \texttt{$\langle$H,H,HadamardGCN,HadamardGCN,min,VEC$\rangle$} & \texttt{$\langle$H,H,Mat,Mat,multiply,VEC$\rangle$}      \\
		\bf     Yelp     &   \texttt{$\langle$H,H,SGC,SGC,multiply,MLP$\rangle$}   &  \texttt{$\langle$H,H,SGC,SGC,multiply,VEC$\rangle$}  &      \texttt{$\langle$H,H,MLP,MLP,multiply,VEC$\rangle$}        \\
		\bf Amazon-Book & \texttt{$\langle$H,H,SGC,SGC,min,MLP$\rangle$} & \texttt{$\langle$ID,ID,SGC,SGC,multiply,VEC$\rangle$}& \texttt{$\langle$ID,H,Mat,MLP,max,VEC$\rangle$}\\
		\bottomrule
	\end{tabular}
\end{table*}

\subsection{Experimental Settings}

\subsubsection{Datasets}
We use MovieLens-100K, MovieLens-1M, Yelp, and Amazon-Book for CF tasks. 
The detailed statistics and preprocess stage of datasets are shown in Table~\ref{tab::dataset_details} in Appendix~\ref{sec::appen::dataset}.

\subsubsection{Evaluation metrics}
As for evaluation metrics, 
we choose two widely used metrics for CF tasks, Recall$@K$ and NDCG$@K$.
According to recent works~\cite{he2017neural, wang2019neural}, 
we set the length for recommended candidates $K$ as 20. 
We use Recall@20 and NDCG@20 as validation.
As for loss function, 
we use BPR loss~\cite{rendle2009bpr}, 
the state-of-the-art loss function for optimizing recommendation models.

\subsubsection{Baselines for Comparison}
\label{sec::experiments::baselines}

Since we encode both hyperparameters and architectures, we can use previous search algorithms on hyperparameters~\cite{bergstra2012random, snoek2012practical, falkner2018bohb, tiao2021bore} and extend them on a joint search space. The details of search algorithms can be found in Appendix~\ref{sec::append::baseline_algo}.

\subsection{Performance Comparison (RQ1)}
For CF tasks, we compare our results with NN-based CF models and Graph-based CF models.
Besides, we also compare our search algorithm with other search algorithms designed for CF models.
The search space is based on analysis of hyperparameter understanding in Section~\ref{sec::strategy::screen_hp}.
We report the performance on four datasets in Table~\ref{tab::model_perform}. 

We summarize the following observation:

\begin{itemize}[leftmargin=*]
	\item 
	We find that in our experiment, our CF model trained by searched hyperparameters and architectures can achieve better performance than the classical CF models. 
	Some single models also perform well due to their special design. 
	For example, LightGCN performs well on ML-100K and Yelp, while NGCF also performs well on ML-1M.
	Since our search algorithm has included various operations used in CF architectures, the overall architecture space can cover the architectures of classical models. 
	\item 
	Compared to NAS method on CF models, our method works better than SIF for considering multiple operations in different stages of architecture. 
	Our methods also outperform AutoCF by 3.10\% to 12.1\%.
	The reason for that is we have an extended joint search space for both hyperparameters and architectures.
	Besides, our choice for hyperparameters is shrunk to a proper range to improve search efficiency and performance.
	\item 
	We can observe in Table~\ref{tab::model_perform} that our searched models can achieve the best performance compared with all other baselines. Note that our proposed method can outperform the best baseline by 2.33\% to 13.02\%.
	The performance improvement on small datasets is better than that on large ones.
\end{itemize}

\subsection{Algorithm Efficiency (RQ2)}

\begin{figure*}[h]
	\centering
	\begin{subfigure}[t]{0.24\textwidth} 
		\centering
		\includegraphics[width=\linewidth]{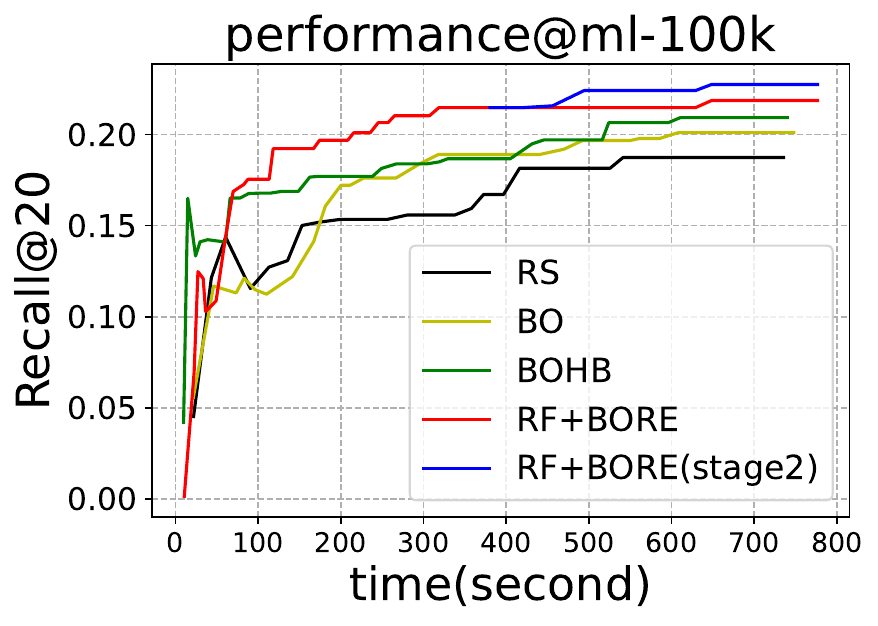} % 
%		\vspace{-15px}
		\caption{}
		\label{fig:search_algo_compare}
	\end{subfigure}
	\begin{subfigure}[t]{0.24\textwidth} 
		\centering
		\includegraphics[width=\linewidth]{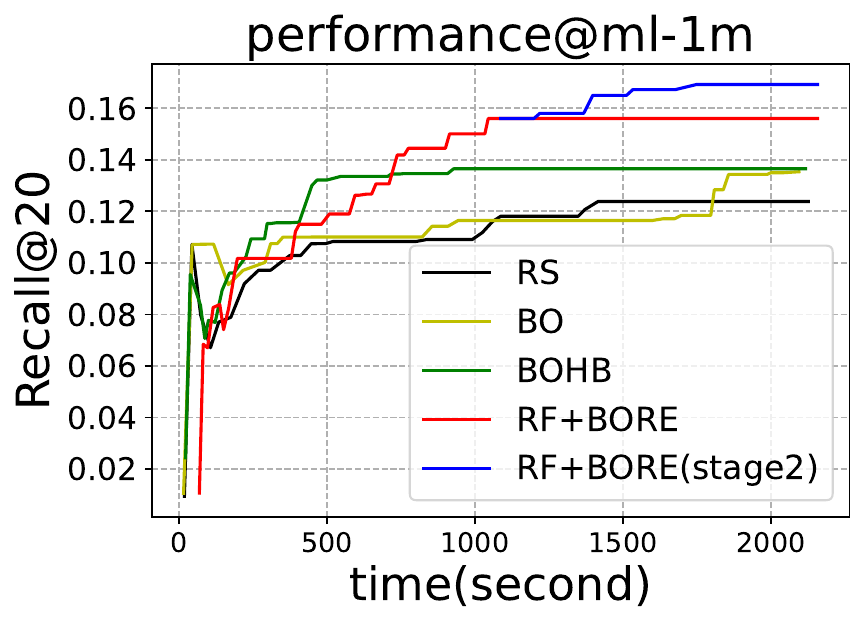} % 
%		\vspace{-15px}
		\caption{}
		\label{fig:search_algo_compare_ml_1m}
	\end{subfigure}
	\begin{subfigure}[t]{0.24\textwidth}
		\centering
		\includegraphics[width=\linewidth]{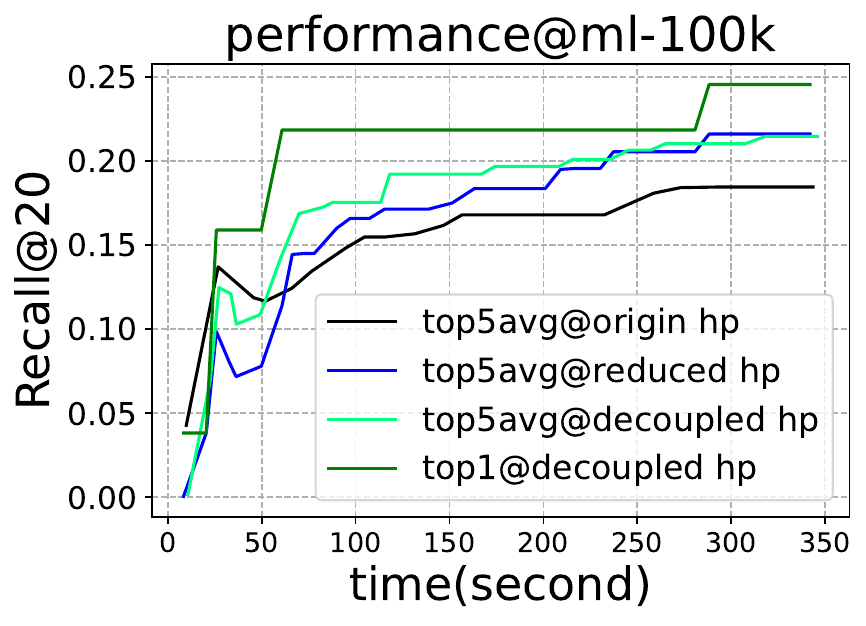} % 
%		\vspace{-15px}
		\caption{}
		\label{fig:hp_ablation}
	\end{subfigure}
	\begin{subfigure}[t]{0.24\textwidth}
		\centering
		\includegraphics[width=\linewidth]{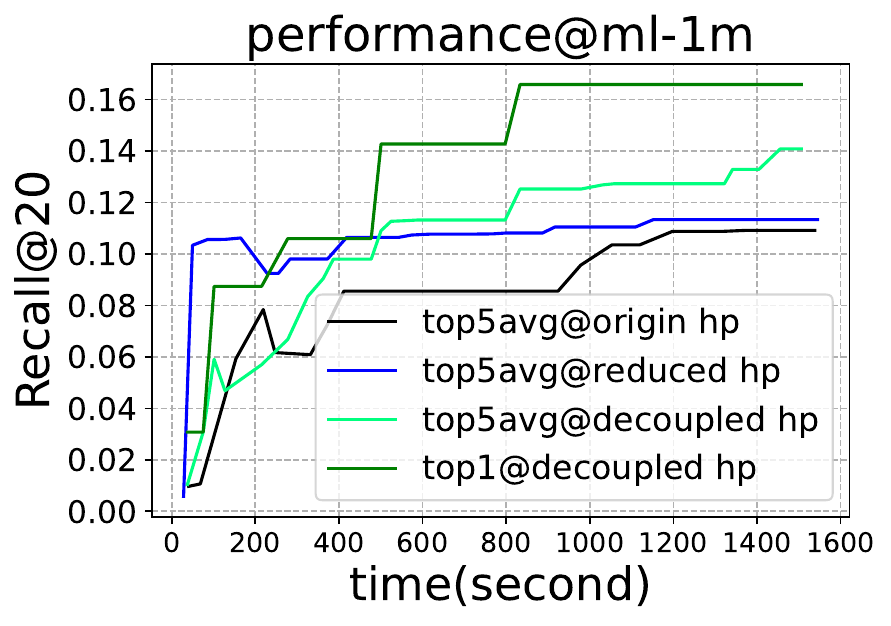} % 
%		\vspace{-15px}
		\caption{}
		\label{fig:hp_ablation_ml1m}
	\end{subfigure}
%	\vspace{-10pt}
	\caption{Comparison for different search algorithms and HP space. 
		(a)-(b) Time curve of different search algorithms on ML-100K and ML-1M; 
		(c)-(d) ablation studies on our search strategy in different search space on ML-100K and ML-1M.}
	\label{fig:perf_ablation}
\end{figure*}

We compare the different search algorithms mentioned in Section~\ref{sec::experiments::baselines}. 
The search results are shown in Figure~\ref{fig:search_algo_compare} on dataset ML-100K and Figure~\ref{fig:search_algo_compare_ml_1m} on dataset ML-1M. 
We plot our results by the output of the search algorithm.
As is demonstrated in Figure~\ref{fig:search_algo_compare_ml_1m} and~\ref{fig:search_algo_compare}, 
search algorithms of BO perform better than random search, since the BO method considers a Gaussian Process surrogate model for simulating the relationship between performance output and hyperparameters and architectures. 
We find that BORE+RF outperforms other search strategies in efficiency. 
The reason is that the surrogate model RF can better classify one-hot encoding of architectures. 
Besides we also compare different time curves for BORE+RF in single-stage and two-stage evaluations. 
%Our two-stage algorithm used a subsampling method on the rating matrix, which can improve search efficiency. 
%Since the surrogate model can be transferred to the origin dataset with high consistency in performance, it can achieve better performance.
Since our two-stage algorithm uses subsampling method on rating matrix, and ensure the consistency between subgraph and whole graph, we can achieve better performance and higher search efficiency.

\subsection{Ablation Study (RQ3)}
\label{sec::search_ablation}
In this subsection, we analyze how important and sensitive the various components of our framework are.
We focus on the performance's improvement and efficiency by reducing and decoupling the hyperparameter space. 
We also elaborate on how we choose sample ratio and effectiveness of tuning hyperparameters.

\subsubsection{Plausibility of screening hyperparameters}
To further validate the effectiveness of our design of screening hyperparameter choices, we choose hyperparameters on the origin space, shrunk space, and decoupled space.
The search time curve is shown in Figure~\ref{fig:hp_ablation}.

We demonstrate that screening hyperparameter choices can help improve performance on CF tasks and search efficiency. According to our results, performance on $\hat{\mathcal{H}}$ is better than the one on the origin space $\mathcal{H}$.
The reason is that the shrunk hyperparameter space has a more significant possibility of including better HP choices for CF architectures to achieve better performance. 
The search efficiency is improved since the choice of hyperparameters is reduced, and proper hyperparameters can be found more quickly in a smaller search space. 
We also find that the search efficiency reduces when we increase the batch size and embedding dimension when we search on the decoupled search space. 
While increasing batch size and embedding dimension help improve final performance, the cost evaluation is higher, which may reduce search efficiency.

\subsubsection{Choice of Sampling Ratio}
To find the impact of changing sampling ratio, we search with different sampling ratios on different datasets in the same controlled search time.  
The evaluation results (Recall@20) for experiments on different sampling ratio settings are listed in Table~\ref{tab::subsample_ratio}.

According to the table, smaller sampling ratios have lower performance. 
The reason is that the user-item matrix sampled by low sample ratio may not capture the consistency in the origin matrix,  as the consistency results shown in Section~\ref{sec::strategy::arch_transfer}.
While sampled dataset generated by higher sample ratio has more considerable consistency with the original dataset, the results in a limited time may not be better. 
The reason is that too much time on evaluation in the first stage may have fewer results for the surrogate model to learn the connection between performance and configurations of hyperparameters and architectures.
Thus, the first stage has a trade-off between sample ratio and time cost,  and we choose 20\% as our sample ratio in our experiments.

\begin{table}[t]
%	\small
	\centering
	\caption{Performance with different sample ratio $\gamma$.}
	\label{tab::subsample_ratio}
%	\vspace{-10pt}
	\begin{tabular}{c|cccc}
		\toprule
		\bf  sample ratio & 5\%  & 10\% & \bf 20\% & 50\%  \\
		\midrule
	\bf	ML-100K    	&  0.2113 	& 0.2224 	&  \bf 0.2646 	&  0.2623  \\
	\bf	ML-1M     	&  0.1471  	& 0.1632	&  \bf 0.1787	&  0.1715  \\
	\bf	Yelp     	&  0.0523  	& 0.0654 	&  \bf 0.0721 	&  0.0706  \\
	\bf	Amazon-Book	&  0.0311  	& 0.0342 	& \bf 0.0365 	&  0.0356   \\
		\bottomrule
	\end{tabular}
\end{table}

\subsubsection{Tuning on Hyperparameters}

To show the effectiveness of our design on joint search, we propose to apply our joint search method to previous research work AutoCF~\cite{gao2021efficient}.
The results are shown in Table~\ref{tab::hp_tune_compare}. 
We apply our search strategy to AutoCF to include hyperparameters in our search space,  noted as AutoCF(HP).
We find that hyperparameter searched on shrunk space can perform better than the one that uses architecture space and random search on different datasets. 

\begin{table}[t]
%	\small
	\centering
	\caption{Comparison on search algorithm with/without HP tuning.}
	\label{tab::hp_tune_compare}
%	\vspace{-10pt}
	\begin{tabular}{c|cccc}
		\toprule
		\bf  Performance & AutoCF & AutoCF (HP) & Improvement  \\
		\midrule
		\bf	ML-100K    	&  0.2259 	& 0.2335 	&   3.36\%	 \\
		\bf	ML-1M     	&  0.1309  	& 0.1358 	&   3.74\% 	\\
		\bf	Yelp     	&  0.0643  	& 0.0672 	&   4.51\%    \\
		\bf	Amazon-Book	&  0.0354  	& 0.0359 	&  	1.41\%   \\
		\bottomrule
	\end{tabular}
\end{table}

\subsection{Case Study (RQ4)}
In this part, we mainly focus on our search strategy's search results of different architectures. 
The search results with top performance share some similarities, while different architecture operations have different performances.

According to search results, we present the top models for each task on all datasets in Table~\ref{tab::case_study_top}.
It is easy to find that interaction history-based encoding features may have better performance and more powerful representative ability. Besides, the embedding function of \texttt{SGC} has stronger representative ability since it can capture the high-order relationship. 
Both \texttt{SGC} and \texttt{HadanardGCN} collect infomation from different layers, simply designed \texttt{SGC} have stronger ability.
As for the interaction function, we find the classical element-wise \texttt{multiply} can receive strong performance; The prediction function of learnable vector and \texttt{MLP} can capture more powerful and complex interaction than \texttt{SUM}.
We can find that these top models of each task have similar implementations, but there also have some differences among different datasets. One top architecture on a given dataset may not get the best performance on another one.

In summary, the proper architectures for different datasets may not be the same, 
but these top results may share some same operations in architecture structures. 
The result and analysis can help human experts to design more powerful CF architectures.

\section{Conclusion and future work}
\label{sec::conclusion}
In this work, we consider a joint search problem on hyperparameters and architectures for Collaborative Filtering models. 
We propose a search framework based on a search space consisting of frequently used hyperparameters and operations for architectures. 
We make a complete understanding of hyperparameter space to screen choices of hyperparameters, 
We propose a two-stage search algorithm to find proper hyperparameters and architectures configurations efficiently. 
We design a surrogate model that can jointly update CF architectures and hyperparameters and can be transferred from small to large datasets.
We do experiments on several datasets, including comparison on different models, search algorithms and ablation study.

For future work, we find it important to model CF models based on Knowledge Graphs as a search problem. 
With additional entities for items and users, deep relationships can be mined for better performance. 
An extended search framework can be built on larger network settings.
Models on extensive recommendation tasks and other data mining tasks can also be considered as a search problem.

\begin{acks}
This work is partially supported by the National Key Research and Development Program of China under 2021ZD0110303, the National Natural Science Foundation of China under 62272262, 61972223, U1936217, and U20B2060, and the Fellowship of China Postdoctoral Science Foundation under 2021TQ0027 and 2022M710006.
\end{acks}

\clearpage

\bibliographystyle{ACM-Reference-Format}
\balance
\bibliography{bibliography}
\clearpage
\nobalance

\appendix
\setcounter{table}{0}   
\setcounter{figure}{0}
\setcounter{algorithm}{0}
\renewcommand{\thetable}{A\arabic{table}}
\renewcommand{\thefigure}{A\arabic{figure}}
\renewcommand{\thealgorithm}{A\arabic{algorithm}}

\section{Appendix}

\subsection{Search Space}
The main notations in this paper are listed in Table~\ref{tab::notations}, and we discuss some details for search space in this section. 

\subsubsection{Hyperparameter Choice}
We explain how to reduce the hyperparameters by Figure~\ref{fig:perf_rank_hp_space} in this section.
We shrink hyperparameter search space based on the performance ranking distribution. 
We can split the hyperparameters into four categories:

\begin{itemize}[leftmargin=*]
	\item Reduction: 
	This kind of hyperparameter is usually categorical, like optimizer. 
	The choices of hyperparameters can be reduced.
	\item Shrunk range: 
	This kind of hyperparameter is usually selected in a continuous range, such as learning rate.
	The choices of these values can be constrained to a smaller range.
	\item Monotonous related:
	The performance with  this kind of hyperparameter usually rises when the hyperparameter increase, such as embedding dimension. However, we can not choose these HPs with too large values for limited memory. Thus, we choose a smaller value in first stage and a larger one in second stage.
	\item No obvious pattern:
	We do not have to change this kind of HP in our experiment, just as weigh decay. 
\end{itemize}

%\subsubsection{Architecture Choice}
%We explain about the choices for embedding functions in this section.
%
%As for GNN-based search operation, 
%\texttt{GraphSAGE}~\cite{hamilton2017inductive} first samples the user-item graph by importance of neighborhoods,
%then concatenates the vectors from sampled graphs as embedding for users/items;
%\texttt{HadamardGCN}~\cite{wang2019neural} constructs messages not only from one-step neighborhoods,
%but also from the user/item node itself;
%\texttt{SGC}~\cite{he2020lightgcn} is an easily designed GCN, combining output from each layer and adding learnable weights between different layers.

\subsection{Search Algorithms}

\subsubsection{Surrogate Model Design}
We demonstrate our search algorithm with surrogate model in Algorithm~\ref{alg::surrogate}.

We design our search algorithm with BORE and Random Forest (RF) regressor.  
%It consists of a surrogate model with an regressor. 
In the first stage, we train the surrogate model with $D$, update parameters of surrogate model.  
The output of BORE+RF can help give an inference between $(0,1)$, and we choose the least one as output. 
After the first stage, we save the parameters of this surrogate model, and transfer it to second stage, which we do experiment on larger datasets.
The configuration of hyperparameters and architectures and the performance on larger dataset can also update the parameters of surrogate model. 
With the knowledge we learn on the first stage, the surrogate model can better choose the proper architecture and hyperparameter for CF tasks.

\begin{algorithm}[t]
	\caption{\texttt{RFSurrogate} RF+BORE surrogate model}
	\begin{flushleft}
		\label{alg::surrogate}
		\textbf{Input}: Training data $S_{\textit{tra}}$ and evaluation data $S_{\textit{val}}$, 
		Reduced hyperparameter space $\hat{\mathcal{H}}$,  
		Architecture space $\mathcal{A}$ , percentage threshold $\tau=0.2$, RF regressor $y=c(\alpha||h;\theta)$ \\
		\textbf{Output}: A proper configuration of hyperparameters $h^*$ and architecture $\alpha^*$\\
	\end{flushleft}
	
	\begin{algorithmic}[1] %
		\STATE Initialize configuration set $D \leftarrow \{\} , i\leftarrow 0$  \\
		\% \textbf{Initialize surrogate model}; \\
		\FOR{i = 1, 2, \dots, N}
		\STATE Select architecture $\alpha_i \in \mathcal{A}$, hyperparameter $h_i\in \mathcal{H}$ ;
		\STATE Get CF evaluation performance $p_i = \mathcal{M}(f(\mathbf{P}^{*}; \alpha_i, h_i), \mathcal{S}_{\textit{val}})$;
		\STATE Save to set $D \leftarrow D \bigcup \{ \alpha_i, h_i, p_i \} $; \\
		\% \textbf{BORE} \\
		\STATE Find $p_\tau$ as the $\tau$-quantile of the performance set $\{p_i\}$ ;
		\IF{$p_i < \tau$}
		\STATE Set $z_i$ label 0;
		\ELSIF{$p_i \geq \tau$}
		\STATE Set $z_i$ label 1;
		\ENDIF
		\STATE Fit the surrogate model with $D$;
		\ENDFOR
		
		\% \textbf{Update surrogate model} (with a new data);\\
		\STATE Get performance $p_k = \mathcal{M}(f(\mathbf{P}^{*}; \alpha_k, h_k), \mathcal{S}_{\textit{val}})$; 
		\STATE Save to set $D \leftarrow D \bigcup \{ \alpha_k, h_k, p_k \} $ ;
		\STATE Set $z_i$;
		\STATE Fit to update RF model $c(;\theta)$ ;
		
		\% \textbf{Find next choice} through $c(\cdot;\theta)$; \\
		\STATE $z_{\text{max}}\leftarrow 0$
		\STATE Randomly sample configuration set  $H=\{h_l, \alpha_l \}$ ;
		\FOR{configuration in $H$}
		\STATE $z_l=c(\alpha_l||h_l;\theta)$;
		\IF{$z_l > z_{\text{max}}$}
		\STATE $h^*$, $\alpha^* = h_l, \alpha_l$;
		\ENDIF
		\ENDFOR

		\STATE return $h^*$, $\alpha^*$
	\end{algorithmic}
\end{algorithm}

\begin{table}[t]
	%	\small
	\centering
	\caption{Notations}
	\label{tab::notations}
	%	\vspace{-10pt}
	\begin{tabular}{c|c}
		\toprule
		\bf  Variable & Definition \\ 
		\midrule
		$h, h^*$	& Hyperparameters (HP), best HP from $\mathcal{H}$ \\
		$\hat{h}$ & HP from $\mathcal{\hat{H}}$\\
		$h^{(i)}$ & The $i$-th HP of $h$, $h=\left(h^{(1)}, h^{(2)}, h^{(i)}, \dots, h^{(n)}\right)$ \\
		$H_i$ & Discrite values set of $h^{(i)}$ \\
		$\lambda, \lambda_1, \lambda_2$ & Some values of hyperparameter from $H_i$ \\
		$\texttt{rank}(\mathsf{h}, \lambda)$ & Relative performance ranking when $h^{(i)}=\lambda$ \\
		$\mathcal{H}$     	&   Origin space of hyperparameters    \\
		$\mathcal{H}_i$     	&   Subset of $\mathcal{H}$, the $i$-th HP is selected from $H_i$ \\ 
		$\mathcal{\hat{H}}$     	&  Subset of $\mathcal{H}$, screened space of hyperparameters    \\
		\midrule
		$\alpha, \alpha^*$ & Architecture/Best architecture from $\mathcal{A}$ \\
		$\mathcal{A}$    	&    Space of architecture     \\
		\midrule
		$\mathcal{S}$    	&    Origin dataset (whole graph)     \\
		$\mathcal{\hat{S}}$ & The $i$-th subsampled dataset (subgraph) \\
		$\gamma$ &  Subsample ratio \\
		$p$, $p_i$ & Test performance \\
		$D$ & Records of $(\alpha, h, p)$  \\
		$c(\cdot; \theta)$ & Surrogate model with parameter $\theta$ \\
		\bottomrule
	\end{tabular}
\end{table}

\subsubsection{Fair Comparison}
Compared with hyperparameters, the choice of architectures can affect the performance of CF models more.
Thus, to compare different experiment settings fairly, we use the average of top5 configurations instead for the search time curve, noted as \texttt{top5avg}.     

\begin{figure}[t]
	\centering
	\includegraphics[width=\linewidth]{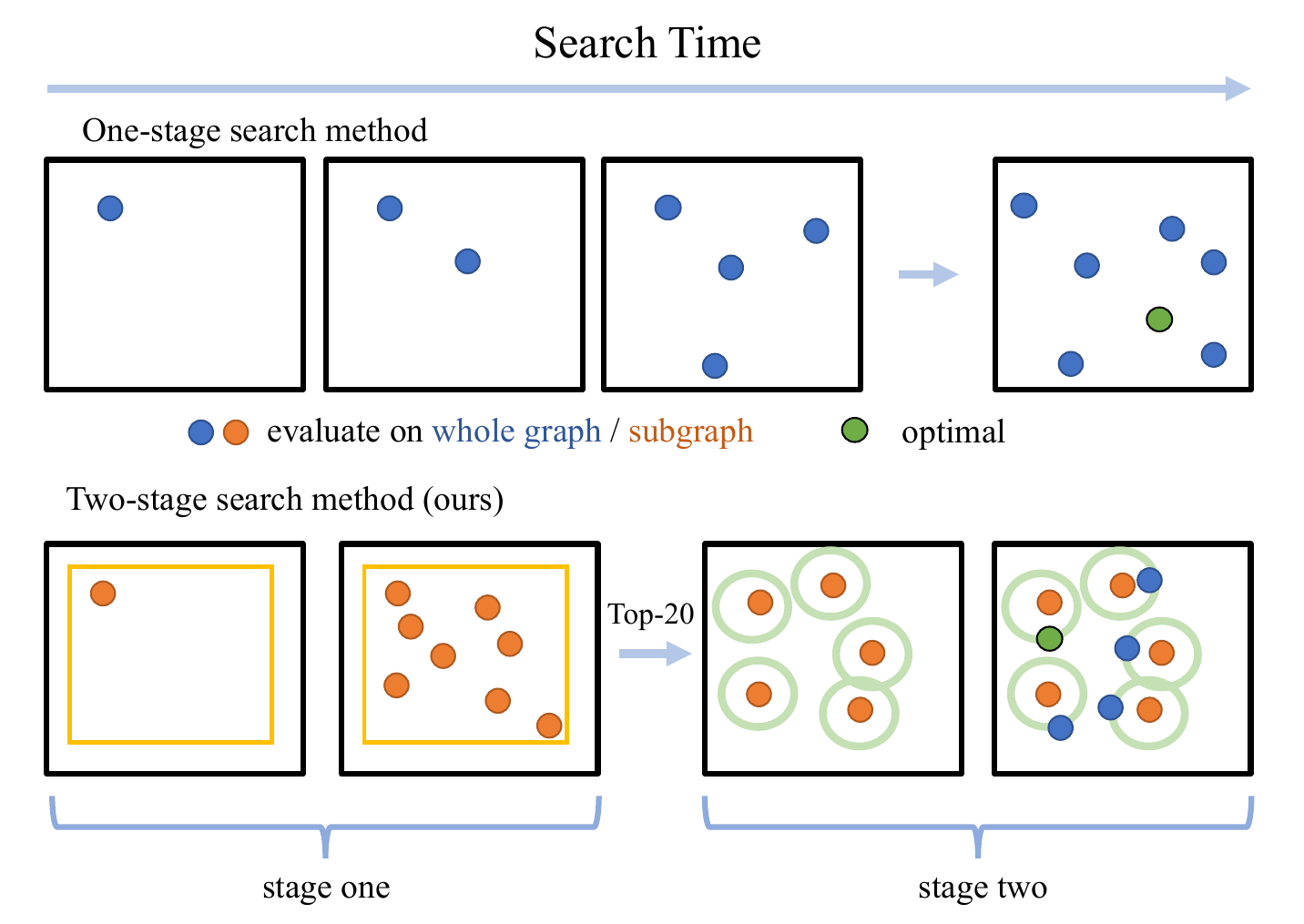}
%	\vspace{-13pt}
	\caption{Comparison of search procedure using one-stage method and our method.}
	\label{fig:search_procedure}
\end{figure}

\subsubsection{Search Procedure}
We show the comparison of conventional joint search method and our method in Figure~\ref{fig:search_procedure}. 
In Figure~\ref{fig:search_procedure}, conventional one-stage  method search components separately, while our method search hyperparameters on a shrunk space. 
Besides, the evaluation time is lower in the first stage in our method.

\subsection{Discussion}
\label{sec::strategy::discussion}

In this section, we discuss about the method we have chosen, and the difference with previous works on joint search problems.

The first is the reason for screening in the hyperparameters space rather than the architecture space.
The search space of hyperparameters mainly consists of components of continuous values with infinite choices. 
And screening range of hyperparameters~\cite{zhang2022knowledge,wang2022profiling} is proven effective on deep neural networks and graph-based models.
Architecture space components are typically categorized, and each one is necessary in some manner and should not be ignored.
Furthermore, we believe that it is unnecessary to reduce the architecture space after conducting a fair ranking of different architectures.

%The second is how our methodology differs from previous research studies on joint search, such as~\cite{dong2020autohas, zela2018towards, seng2023feathers, cai2021stnas}.
Our work is the first on CF tasks, which is different from previous research studies on joint search.
\cite{zela2018towards} mainly focuses on a joint search problem on typical neural networks.
Another study on the joint search problem, AutoHAS~\cite{dong2020autohas}, focuses on model search with weight sharing.
FEATHERS~\cite{seng2023feathers} focuses on a joint search problem on Federate Learning, an aspect of Reinforcement Learning.
In ST-NAS~\cite{cai2021stnas}, the sub-models are candidates of a designed super-net, and they sample sub-ST-models from the super-net and weights are update using training loss while updating HPs with the validation loss.

%As for sampling method, we sample on items but not users because we are motivated to recommend different items for different users through users' preference. Thus, the mostly connected items can be the most representative ones.  
%As for the frequency-base sample algorithm, we can also consider it in a graph perspective, the frequency of an item equals its degree in a bipartite graph. 

\subsection{Data Preprocess}
\label{sec::appen::dataset}

\subsubsection{Details of Datasets}
In our experiments, we choose four real-world raw datasets and build them for evaluation on CF tasks. 

\begin{itemize}[leftmargin=*]
	\item \textbf{MovieLens-100K}
	\footnote{https://grouplens.org/datasets/movielens/100k} 
	This widely used movie-rating dataset
	contains 100,000 ratings on movies from 1 to 5. We also convert
	the rating form to binary form, where each entry is c vmarked as 0
	or 1, indicating whether the user has rated the item.
	\item \textbf{MovieLens-1M}
	\footnote{https://grouplens.org/datasets/movielens/1m}
	This widely used movie-rating dataset
	contains more than 1 million ratings on movies from 1 to 5. Similarly, for	MovieLens-1M, we  build an implicit dataset.
	\item \textbf{Yelp} 
	\footnote{https://www.yelp.com/dataset/download}
	This is published officially by Yelp, a crowd-sourced review
	forum website where users can write comments and reviews
	for various POIs, such as hotels, restaurants, etc.
	\item \textbf{Amazon-Book}
	\footnote{https://nijianmo.github.io/amazon/index.html}
	This book-rating dataset is collected from
	users’ uploaded review and rating records on Amazon.
\end{itemize}

\begin{table}[t]
%	\small
	\centering
	\caption{Statistics of datasets}
	\label{tab::dataset_details}
%	\vspace{-10pt}
	\begin{tabular}{c|cccc}
		\toprule
		\bf  Name & $\#$users & $\#$items &$\#$records & density \\ \midrule
		\bf	ML-100K    	&    943    & 1,682 &  100,000  &  6.304\%  \\
		\bf	ML-1M     	&   6,040  & 3,952 &  1,000,209 &   4.190\%   \\
		\bf	Yelp     	&    6,102  & 18,599 &  445,576   &  0.393\%  \\
		\bf	Amazon-Book &  25,774  & 80,211 & 3,040,864  &  0.147\%   \\
		\bottomrule
	\end{tabular}
\end{table}

\subsubsection{Preprocess}
Since the origin dataset in Table~\ref{tab::dataset_details} is too large for training, 
we reduce them in frequency order. 
We use 10-core select on Yelp and 50-core select on Amazon-Book.
$N$-core means we choose users and items that appear more than $N$ times in the whole record history.
After we select the dataset, we split the data into training, validation and test sets. 
We shuffle each set when we start a new evaluation task.

\subsubsection{Dataset sampling}
Some papers studying long-tail recommendation~\cite{zheng2021disentangling} or self-supervised recommendation~\cite{wu2021self} may discuss the impact of data sparsity.
We can sample the subgraph according to the popularity~\cite{he2016fast}, mainly user-side and item-side, and the long-tail effect on the item-side is more severe.
Thus we sample the rating matrix based on the item frequency. 
The more the item appears in rating records, the more likely it is reserved in the subsampled matrix.

\subsection{Detailed Settings of Experiments}
\subsubsection{Baseline algorithms}
\label{sec::append::baseline_algo}
\begin{itemize}[leftmargin=*]
	\item \textbf{Random Search}~\cite{bergstra2012random}.
	It is a simple method for finding hyperparameters and architectures with random choices on either continuous or categorical space.
	\item \textbf{Bayesian Optimization}~\cite{snoek2012practical}.
	Bayesian Optimization (BO) is a search algorithm based on the analysis of posterior information, and it can use Gaussian Process as a surrogate model. 
	\item \textbf{BOHB}~\cite{falkner2018bohb}.
	BOHB is a method consisting of both Bayesian Optimization (BO) and HyperBand (HB), helping modulate functions with a lower time budget.
	\item \textbf{BORE}~\cite{tiao2021bore}.
	BORE is a BO method considering expected improvement as a binary classification problem. 
	It can be combined with regressors, including Random Forest (RF), Multi-Layer Perceptron (MLP), and Gaussian Process (GP). We choose RF as a regressor in our experiments.
\end{itemize}

\subsubsection{Hardware Environment}
We implement models with PyTorch 1.12 and run experiments on a 64-core Ubuntu 20.04 server with NVIDIA GeForce RTX 3090 GPU with 24 GB memories each. 
It takes 3.5-4 hours to search on a dataset with one million records.
\subsubsection{Codes}
%Our coding files can be found in~\href{https://anonymous.4open.science/r/CodeForKDD23-17BE}{HERE}.
We have released our implementation code for experiments in~\href{https://github.com/overwenyan/Joint-Search}{https://github.com/overwenyan/Joint-Search}

%\href{https://github.com/overwenyan/CodeForKDD23}{HERE}.

\end{document}